%% file: paper.tex
\documentclass{sty/jpaper}
\input{sty/mysty.tex}

\newif\ifcameraready
\camerareadytrue
\newcommand{\versionnum}[0]{4.1}

\ifcameraready
\else
\fi

\ifcameraready
  \newcommand{\todo}[1][]{}
  \newcommand{\ch}[0]{}
\else
  \newcommand{\todo}[1][]{\textbf{\fcolorbox{black}{red}{\color{white}{TODO}}} \underline{$\overline{\hbox{\emph{#1}}}$}}
  \newcommand{\ch}[1]{{\color{BrickRed} #1}}
\fi

\title{Adaptive-Latency DRAM:\\Reducing DRAM Latency by Exploiting Timing Margins}

\author{
	Donghyuk Lee$^{1,2}$\qquad
 	Yoongu Kim$^{2}$\qquad
 	Gennady Pekhimenko$^{3,2}$ 
	\vspace{2pt}\\
 	Samira Khan$^{4,2}$\qquad
 	Vivek Seshadri$^{5,2}$\qquad
 	Kevin Chang$^{6,2}$\qquad
 	Onur Mutlu$^{7,2}$}
\affil{
 	$^1${\em NVIDIA Research}\qquad
 	$^2${\em Carnegie Mellon University}\qquad
 	$^3${\em University of Toronto}\vspace{2pt}\\
 	$^4${\em University of Virginia}\qquad
 	$^5$\large{\em Microsoft Research India}\qquad
 	$^6${\em Facebook}\qquad
 	$^7${\em ETH Z{\"u}rich}
}

\begin{document} 
	\date{}
	\maketitle

	\input{sec/0_abstract}
	\input{sec/1_summary}

	\input{sec/1a_relatedwork}
	\input{sec/2_significance}

        \input{sec/2a_conclusion}
	\input{sec/3_acknowledge}

{
\setstretch{0.96}
	\bibliographystyle{sty/IEEEtranS}
	\bibliography{ref/paper}
}

\end{document}

%% file: sty/mysty.tex
\usepackage[nocompress]{cite}
\usepackage{algorithmic}
\usepackage{array}

\usepackage{fixltx2e}
\usepackage{dblfloatfix}
\usepackage[nolessnomore, italic]{mathastext}
\usepackage[T1]{fontenc}
\usepackage[usenames,dvipsnames,svgnames,table]{xcolor}
\usepackage[normalem]{ulem}
\usepackage{enumitem}
\usepackage{setspace}
\usepackage{indentfirst}
\usepackage{footmisc}
\usepackage{fancyhdr}
\usepackage{authblk}
\usepackage[binary-units=true]{siunitx}
\usepackage[us,12hr]{datetime}
\usepackage[keeplastbox]{flushend}
\usepackage[hidelinks]{hyperref}

\usepackage{url}
\usepackage{xspace}
\usepackage{graphicx}
\usepackage{wrapfig}
\usepackage{booktabs}
\usepackage{multirow}
\usepackage{amsmath}
\usepackage{amssymb}
\usepackage{stmaryrd}
\usepackage{indentfirst}
\usepackage{footnote}
\usepackage{float}
\usepackage{subfig}
\usepackage{gensymb}

\let\oldcelsius\celsius
\renewcommand{\celsius}{~\oldcelsius\xspace\xspace}
\newcommand{\ALD}[0]{AL-DRAM\xspace}

\newcommand{\squishlist}{
	\begin{list}{$\bullet$}
		{ \setlength{\itemsep}{0pt}      \setlength{\parsep}{3pt}
			\setlength{\topsep}{3pt}       \setlength{\partopsep}{0pt}
			\setlength{\leftmargin}{1.5em} \setlength{\labelwidth}{1em}
			\setlength{\labelsep}{0.5em} } }
\newcommand{\squishend}{
    \end{list}  }

\newcommand{\DIMMs}[0]{115\xspace}

\newcommand{\ReadHot}[0]{21.1}
\newcommand{\WriteHot}[0]{34.4}
\newcommand{\trcdHot}[0]{15.6}
\newcommand{\trasHot}[0]{20.4}
\newcommand{\twrHot}[0]{20.6}
\newcommand{\trpHot}[0]{28.5}
\newcommand{\ReadCold}[0]{32.7}
\newcommand{\WriteCold}[0]{55.1}
\newcommand{\trcdCold}[0]{17.3}
\newcommand{\trasCold}[0]{37.7}
\newcommand{\twrCold}[0]{54.8}
\newcommand{\trpCold}[0]{35.2}

%% file: sec/0_abstract.tex
\begin{abstract}

This paper summarizes the idea of Adaptive-Latency DRAM (AL-DRAM), 
which was published in HPCA 2015~\cite{lee-hpca2015}, and examines
the work's significance and future potential. 
AL-DRAM is a mechanism
that optimizes DRAM latency based on the DRAM module and the operating
temperature, by exploiting the extra margin that is built into the DRAM timing
parameters. 
DRAM manufacturers provide a large margin for the timing parameters as a
provision against two worst-case scenarios.  First, due to process variation,
some outlier DRAM chips are much slower than others.
Second, chips become slower at higher temperatures.
The timing parameter margin ensures that the slow outlier chips operate
reliably at the worst-case temperature, and hence leads to a high access latency.

% The key observation is that the timing
% parameters are dictated by the worst-case temperatures and worst-case DRAM
% cells, both of which lead to a small amount of charge storage and hence high
% access latency. One can therefore reduce the DRAM latency by adapting the timing
% parameters to the current operating temperature and the current DRAM module that
% is being accessed. 

Using an FPGA-based DRAM testing
platform, our work first characterizes the extra margin
for \DIMMs DRAM modules from three major manufacturers. The experimental results
demonstrate that it is possible to reduce four of the most critical timing
parameters by a minimum/maximum of \trcdCold\%/\twrCold\% at 55\celsius~while
maintaining reliable operation. 
AL-DRAM uses these observations to adaptively select reliable DRAM timing parameters 
for each DRAM module based on the module's current operating conditions.
AL-DRAM does not require any changes to the DRAM chip or its
interface; it only requires multiple different timing parameters to be specified
and supported by the memory controller. Our real system evaluations show that
AL-DRAM improves the performance of memory-intensive workloads by an average of
14\% without introducing any errors.

Our characterization and proposed techniques have inspired several other works
on analyzing and/or exploiting 
\ch{different sources of latency and performance variation within DRAM
chips}~\cite{chang-sigmetrics2016, chang-sigmetrics2017, lee-sigmetrics2017,
kim-hpca2018, hassan-hpca2016, patel-isca2017}.

\end{abstract}

%% file: sec/1_summary.tex
\section{Problem: High DRAM Latency} \label{sec:problem}

A DRAM chip is made of capacitor-based cells that represent data in the form of
electrical charge. To store data in a cell, charge is injected, whereas to
retrieve data from a cell, charge is extracted. Such {\em movement of charge}
happens through a wire called {\em bitline}. Due to the large resistance and the
large capacitance of the bitline, it takes a long time to access DRAM cells. To
guarantee correct operation for every module sold, DRAM manufacturers impose a set of minimum latency
restrictions on DRAM accesses, called {\em timing parameters}~\cite{jedec-ddr3}.
Ideally, timing parameters should provide {\em just enough} time for a DRAM chip
to operate correctly. In practice, however, there is a very large margin in the
timing parameters to ensure correct operation under {\em worst-case} conditions
with respect to two aspects. First, due to {\em process variation}, some outlier
cells suffer from a larger RC-delay than other cells~\cite{lee-iedm1996,
kang-memforum2014}, and require more time to be accessed. Second, due to {\em
temperature dependence}, DRAM cells lose more charge at high
temperature~\cite{yaney-iedm1987, liu-isca2013}, and therefore require more time
to be accessed. Due to the worst-case provisioning of the fixed timing
parameters, which ensure reliable operation up to a temperature of 85\celsius, it takes a longer time to
access most of DRAM under most operating conditions than is actually necessary for correct
operation.

\section{Key Observations and Our Goal} \label{sec:observation}

First, we observe that \textbf{most DRAM chips do {\em not} contain the worst-case cells that require the largest access
latency.} Using an FPGA-based testing platform~\cite{hassan-hpca2017}, we
profile \DIMMs real DRAM modules and observe that the slowest cell (i.e., the cell that stores the
smallest amount of charge) for a typical chip is still significantly faster than the slowest cell of the
worst-case chip. Our profiling exposes the large margin built into {\em DRAM} timing
parameters. In particular, we identify four timing parameters that are the most
critical during a DRAM access: {\tt tRCD}, {\tt tRAS}, {\tt tWR}, and {\tt
tRP}.\footnote{For a detailed background on the operation of DRAM, and an explanation of each timing
parameter, we refer the reader to our prior works~\cite{kim-isca2012,
kim-micro2010, kim-hpca2010, lee-hpca2013, lee-hpca2015, chang-hpca2014,
chang-hpca2016, chang-sigmetrics2016, chang-sigmetrics2017, hassan-hpca2016,
hassan-hpca2017, khan-cal2016, khan-dsn2016, khan-sigmetrics2014, khan-hpca2014,
liu-isca2012, liu-isca2013, patel-isca2017, seshadri-micro2013,
seshadri-micro2017, lee-sigmetrics2017, kim-cal2015, lee-pact2015, lee-taco2016, kim-isca2014, kim-hpca2018}.} At 55\celsius, we demonstrate that the parameters can be
reduced by an average of \trcdCold\%, \trasCold\%, \twrCold\%, and \trpCold\%, respectively,
while still maintaining correctness.

Second, we observe that \textbf{most DRAM chips are {\em not} exposed to the worst-case temperature of
85\celsius.} We measure the DRAM ambient temperature in a server cluster running
a very memory-intensive benchmark, and find that the temperature {\em never}
exceeds 34\celsius, and never changes by more than
0.1\celsius\xspace per second. Other works~\cite{elsayed-sigmetrics2012,
liu-hpca2011} also observe that worst-case DRAM
temperatures are not common, and that servers typically operate at much lower
temperatures~\cite{elsayed-sigmetrics2012,
liu-hpca2011}.

Based on these two observations, we show {\em that} typical DRAM chips operating at
typical temperatures (e.g., 55\celsius) are capable of operating correctly when
accessed with a much smaller access latency, but are nevertheless forced to
operate at the largest latency of the worst-case module and operating conditions.
Modules in existing systems use these worst-case latencies because existing
memory controllers are equipped with only a single set of timing parameters
that are dictated by the worst case.

{\bf Our goal} in our HPCA 2015 paper~\cite{lee-hpca2015} is to exploit the
extra margin that is built into the DRAM timing parameters to reduce DRAM
latency, and thus improve performance as well as energy consumption. To this end,
we first provide a detailed analysis of {\em why} we can reduce DRAM timing
parameters without sacrificing reliability.

\section{Charge \& Latency Interdependence}
\label{sec:dependence}

The operation of a DRAM cell is governed by two important parameters: {\em
i)}~the quantity of charge and {\em ii)}~the latency it takes to move charge.
These two parameters are closely related to each other. Based on SPICE
simulations with a detailed DRAM model, we identify the quantitative
relationship between charge and latency~\cite{lee-hpca2015}.
We briefly summarize our three key observations from these analyses here.
Section~7 of our HPCA 2015 paper~\cite{lee-hpca2015} provides a detailed
analysis of our observations.

First, having more charge in a DRAM
cell accelerates the {\em sensing} operation in the cell, especially at the
beginning of sensing, enabling the opportunity to shorten the
timing parameters that correspond to sensing ({\tt tRCD} and {\tt tRAS}). Second, when {\em restoring} the
charge in a DRAM cell, a large amount of the time is spent on injecting the
final small amount of charge into the cell. If there is already enough charge
in the cell for the next access, the cell does {\em not} need to be fully
restored. In this case, it is possible to shorten the latter part of the
restoration time, creating the opportunity to shorten the timing
parameters that correspond to restoration ({\tt tRAS} and {\tt tWR}). Third, at the end of {\em precharging},
i.e., setting the bitline into the initial voltage level (before accessing a
cell) for the next access, a large amount of the time is spent on precharging
the final small amount of bitline voltage difference from the initial level.
When there is already enough charge in the cell to overcome the voltage
difference in the bitline, the bitline does {\em not} need to be fully
precharged. Thus, it is possible to shorten the final part of the precharge
time, creating the opportunity to shorten the timing parameter that corresponds to precharge
({\tt tRP}). Based on these three observations, we conclude that {\em timing
parameters can be shortened if DRAM cells have enough charge}.

\section{Adaptive-Latency DRAM} \label{sec:observation}

As explained in Section~\ref{sec:dependence}, the amount of charge in the cell right before an access to it
plays a critical role in how long it takes to retrieve the correct data from the
cell. In Figure~\ref{fig:aldram}, we illustrate the impact of process variation
using two different cells: one is a {\em typical} cell (left column) and the
other is the \emph{worst-case} cell that deviates the most from the typical (right
column). The worst-case cell initially contains less charge than the typical cell
for two reasons. First, due to its {\em large
resistance}, the worst-case cell cannot allow charge to flow inside quickly.
Second, due to its {\em small capacitance}, the worst-case cell cannot store
much charge even when it is fully charged. To accommodate such a worst-case
cell, existing timing parameters are conservatively set to large values. 

\input{fig/worst}

In Figure~\ref{fig:aldram}, we also illustrate the impact of temperature
dependence using two cells at two different operating temperatures: {\em i)} a typical
temperature (55\celsius, bottom row), and {\em ii)} the worst-case temperature
(85\celsius, top row) supported by DRAM standards. Both typical and worst-case
cells leak charge at a faster rate at the worst-case temperature. Therefore, not
only does the worst-case cell have less charge to begin with, but it is left
with {\em even less} charge at the worst temperature because it leaks charge at
a faster rate (top-right in Figure~\ref{fig:aldram}). To accommodate the
combined effect of process variation {\em and} temperature dependence, existing
timing parameters are set to very large values. That is why the worst-case
condition for correctness is specified by the top-right of
Figure~\ref{fig:aldram}, which shows the least amount of charge stored in {\em
the worst-case cell at the worst-case temperature} in its initial state. On top
of this, DRAM manufacturers add an extra latency margin to the access time under
worst-case conditions. In other words, the amount of charge in a cell under worst-case
conditions is still greater than the minimum amount of charge required for correctness.

If we were to reduce the timing parameters, we would also reduce the amount of charge
stored in the cells. It is important to note, however, that we are proposing to
exploit {\em only} the {\em additional slack} (in terms of charge) compared to
the worst case. This allows us to provide as strong of a reliability guarantee
as manufacturers currently do for worst-case cells and operating conditions. 
In Figure~\ref{fig:aldram}, we illustrate the impact of
reducing the timing parameters. The lightened portions inside the cells
represent the amount of charge that we are giving up by using reduced timing
parameters. Note that we are not giving up any charge for the worst-case cell at
the worst-case temperature. Although the other three cells are {\em not} fully
charged in their initial state, w propose to give up just enough charge from them
such that they are left with a similar amount of charge as
the worst case (top-right). This is because these cells are capable of either
holding more charge to begin with (typical cell, left column) or holding their
charge for longer (typical temperature, bottom row). Therefore, optimizing the
timing parameters (based on the amount of existing charge slack) provides the
opportunity to reduce overall DRAM latency while still maintaining the same
reliability guarantees provided by DRAM manufacturers.

Based on these observations, we propose Adaptive-Latency DRAM (\ALD), a
mechanism that dynamically optimizes the timing parameters for different modules
at different temperatures. \ALD exploits the {\em additional charge slack}
present in the common-case compared to the worst-case, thereby preserving the
level of reliability (at least as high as the worst-case) provided by DRAM
manufacturers.

\section{DRAM Latency Profiling:\\ Experimental Analysis of 115 Modules}
\label{sec:profiling}

We present and analyze the results of our DRAM profiling experiments, performed
on our FPGA-based DRAM testing infrastructure, SoftMC~\cite{hassan-hpca2017},
which is also used in our various past works analyzing various DRAM
characteristics~\cite{liu-isca2013, khan-sigmetrics2014, kim-isca2014,
qureshi-dsn2015, lee-hpca2015, chang-sigmetrics2016, khan-dsn2016,
chang-sigmetrics2017, lee-sigmetrics2017}. In total, we analyze 115 DRAM modules
from three major manufacturers, comprising 920 total DRAM chips. Our full
methodology is explained in Section 6 of our HPCA 2015
paper~\cite{lee-hpca2015}.

\subsection{Analysis of a Representative DRAM Module}

We study the possible timing parameter reductions of a DRAM module while still
maintaining correctness. To guarantee reliable DRAM operation, DRAM
manufacturers provide a built-in {\em safety margin} in retention time, also
referred to as {\em a guardband}~\cite{wang-ats2001, ahn-asscc2006,
khan-sigmetrics2014, patel-isca2017, liu-isca2013}. This way, DRAM manufacturers
are able to guarantee that even the weakest cell is insured against various
other modes of failure. We first measure the safety margin of a DRAM module by
sweeping the refresh interval at the worst-case operating temperature (85\celsius),
using the standard timing parameters. Figure~\ref{fig:safety_detail} plots the
maximum refresh intervals of each bank and each chip in a DRAM module for both
read and write operations. We make several observations. First, the maximum
error-free refresh intervals of both read and write operations are much larger
than the DRAM standard (208~ms for the read and 160~ms for the write operations
vs.\ the 64~ms standard). Second, for the smaller architectural units (banks and
chips in the DRAM module), some of them operate {\em without} incurring errors
even at much higher refresh intervals than others (as high as 352~ms for the
read operations and 256~ms for the write operations). This is because the error-free retention
time is determined by the worst single cell in each architectural component
(i.e., bank/chip/module).

\input{fig/profile_detail}

Based on this experiment, we define the {\em safe refresh interval} for a DRAM
module as the maximum refresh interval that leads to no errors, minus an
additional margin of 8 ms, which is the increment at which we sweep the refresh
interval. The safe refresh interval for the read and write operations are 200 ms and
152 ms, respectively. We then use the safe refresh intervals to run the tests
with all possible combinations of timing parameters. For each combination, we
run our tests at two temperatures: 85\celsius\xspace and 55\celsius.

Figure~\ref{fig:latency_detail_read} plots the error-free timing parameter
combinations ({\tt tRCD}, {\tt tRAS}, and {\tt tRP}) in the read operation test. For each
combination, there are two stacked bars --- the left bar for the test at
55\celsius\xspace and the right bar for the test at 85\celsius\xspace. Missing
bars indicate that the test (with that timing parameter combination at that
temperature) causes errors. Figure~\ref{fig:latency_detail_write} plots same
data for the write operation test ({\tt tRCD}, {\tt tWR}, and {\tt tRP}).

We make two observations. First, even at the highest temperature of
85\celsius, the DRAM module reliably operates with reduced timing parameters
(24\% reduction for read, and 35\% reduction for write operations). Second, at
the lower temperature of 55\celsius, the potential latency reduction is even
higher (36\% for read, and 47\% for write operations). These latency reductions
are possible {\em while} maintaining the safety margin of the DRAM module. From
these two observations, we conclude that there is significant opportunity to
reduce DRAM timing parameters {\em without compromising reliability}.

\subsection{Analysis of 115 DRAM Modules} \label{sec:multiple_dimm}

We have studied the effect of temperature and the potential to reduce various
timing parameters at different temperatures for a single DRAM module. The same
trends and observations also hold true for all of the other modules we studied.
In this section, we analyze the effect of process variation by studying the
results of our profiling experiments on \DIMMs DIMMs. We also present results
for intra-chip process variation by studying the process variation across
different banks within each DIMM.

Figure~\ref{fig:read_safety} (solid line) plots the highest refresh interval
that leads to correct operation across all cells at 85\celsius\xspace within
{\em each DIMM} for the read operation test. The red dots on top show the highest refresh
interval that leads to correct operation across all cells within {\em each bank}
for all 8 banks. Figure~\ref{fig:write_safety} plots the same data for the write
operation test.

\input{fig/profile_multiple}

We draw two conclusions. First, although there exist a few modules which {\em
just} meet the timing parameters (with a low safety margin), a vast majority of
the modules very comfortably meet the standard timing parameters (with a high
safety margin). This indicates that a majority of the DIMMs have significantly
higher safety margins than the worst-case module {\em even at the
highest-acceptable operating temperature of 85\celsius}. Second, the effect of
process variation is even higher for banks within the same DIMM, explained by
the large spread in the red dots for each DIMM. Since banks within a DIMM can be
accessed independently with different timing parameters, one can potentially
imagine a mechanism that more aggressively reduces timing parameters at a bank
granularity and not just the DIMM granularity. We leave this for future
work.\footnote{Note that our future works~\cite{lee-sigmetrics2017,
chang-sigmetrics2016, chang-sigmetrics2017, lee-thesis2016, chang-thesis2017}
explain this observation of latency heterogeneity within a DRAM chip.}

To study the potential of reducing timing parameters for each DIMM, we sweep all
possible combinations of timing parameters ({\tt tRCD}/{\tt tRAS}/{\tt tWR}/{\tt
tRP}) for all the DIMMs at both the highest acceptable operating temperature
(85\celsius) and a more typical operating temperature (55\celsius). We then
determine the acceptable DRAM timing parameters for each DIMM for both
temperatures while maintaining its safety margin. 

Figures~\ref{fig:read_latency} and~\ref{fig:write_latency} show the results of
this experiment for the DRAM read and DRAM write, respectively. The y-axis plots
the sum of the relevant timing parameters ({\tt tRCD}, {\tt tRAS}, and {\tt tRP}
for the DRAM read and {\tt tRCD}, {\tt tWR}, and {\tt tRP} for the DRAM write).
The solid black line shows the latency sum of the standard timing parameters
(DDR3 DRAM specification). The dotted red line and the dotted blue line show the
most acceptable latency parameters for each DIMM at 85\celsius\xspace and
55\celsius, respectively. The solid red line and blue line show the average
acceptable latency across all DIMMs.

We make two observations. First, even at the highest temperature of 85\celsius,
DIMMs can reliably operate at reduced access latencies: \ReadHot\% on average
for read, and \WriteHot\% on average for write operations. This is a direct
result of the possible reductions in timing parameters {\tt tRCD}/{\tt
tRAS}/{\tt tWR}/ {\tt tRP} --- \trcdHot\%/\trasHot\%/\twrHot\%/\trpHot\% on
average across all the DIMMs.\footnote{Due to space constraints, we present only
the {\em average} potential reduction for each timing parameter. However,
detailed characterization of each DIMM can be found online at the SAFARI
Research Group website~\cite{safari}.} As a result, we conclude that process
variation and lower temperatures enable a significant potential to reduce DRAM
access latencies. Second, we observe that at lower temperatures (e.g.,
55\celsius) the potential for latency reduction is even greater (\ReadCold\% on
average for read, and \WriteCold\% on average for write operations), where the
corresponding reduction in timing parameters {\tt tRCD}/{\tt tRAS}/{\tt tWR}/
{\tt tRP} are \trcdCold\%/\trasCold\%/ \twrCold\%/\trpCold\% on average across
all the DIMMs.

We conclude that existing DRAM modules can be accessed reliably with lower
access latencies, especially at lower temperatures than the worst-case
temperature specified by DRAM manufacturers.

\section{Real-System Evaluation} \label{sec:evaluation}

We evaluate \ALD on a real system that offers dynamic software-based control
over DRAM timing parameters at runtime~\cite{amd-4386, amd-bkdg}. We use the
minimum values of the timing parameters that do {\em not} introduce any errors
at 55\celsius~for any module to determine the latency reduction at 55\celsius.
Thus, the latency is reduced by 27\%/32\%/33\%/18\% for {\tt tRCD}/{\tt
tRAS}/{\tt tWR}/{\tt tRP}, respectively. Our full methodology is described in
Section~8 of our HPCA 2015 paper~\cite{lee-hpca2015}.

Figure~\ref{fig:result_1r1c} shows the performance improvement of reducing the
timing parameters in the evaluated memory system with one rank and one memory
channel at a 55\celsius~operating temperature. We run a variety of different
applications in two different configurations. The first one (single-core) runs
only one thread, and the second one (multi-core) runs multiple
applications/threads. We run each configuration 30 times (only SPEC benchmarks
are executed 3 times due to their large execution times), and present the
average performance improvement across all the runs and their standard deviation
as an error bar. Based on the last-level cache misses per kilo instructions
(MPKI), we categorize our applications into memory-intensive or non-intensive
groups, and report the geometric mean performance improvement across all
applications from each group. 

\input{fig/result}

We draw three key conclusions from Figure~\ref{fig:result_1r1c}. First, \ALD
provides significant performance improvement over the baseline (as high as
20.5\% for the very memory-bandwidth-intensive STREAM
applications~\cite{moscibroda-usenix2007}). Second, when the memory system is
under higher pressure with multi-core/multi-threaded applications, we observe
significantly higher performance (than in the single-core case) across all
applications from our workload pool. Third, as expected, memory-intensive
applications benefit more in performance than non-memory-intensive workloads
(14.0\% vs.~2.9\% on average). We conclude that by reducing the DRAM timing
parameters using AL-DRAM, we can speed up a real system by 10.5\% (on average
across all 35 workloads on the multi-core/multi-thread configuration).

We also conducted reliability stress tests for our mechanism. We ran our
workloads for 33 days without interruption of the lower latencies. We observed no
errors and correct results.

\section{Other Results and Analyses in Our Paper} \label{sec:other_results}

Our HPCA 2015 paper~\cite{lee-hpca2015} includes significant amount of DRAM
latency analyses and system performance evaluations. We refer the reader
to~\cite{lee-hpca2015} for detailed evaluations and analyses.

\squishlist

	\item {\bf Effect of Changing the Refresh Interval on DRAM Latency.} We
	evaluate DRAM latency at different refresh intervals. We observe that
	refreshing DRAM cells more frequently enables more DRAM latency reduction
	(Section~7.1 of our HPCA 2015 paper~\cite{lee-hpca2015}).

	\item {\bf Effect of Reducing Multiple Timing Parameters.} We study the
	potential for reducing multiple timing parameters simultaneously. Our key
	observation is that reducing one timing parameter leads to decreasing the
	opportunity to reduce another timing parameter simultaneously (Section~7.2
	of our HPCA 2015 paper~\cite{lee-hpca2015}).

	\item {\bf Analysis of the Repeatability of Cell Failures.} We perform tests
	for five different scenarios to determine that a cell failure due to reduced
	latency is repeatable: {\em i)} same test, {\em ii)} test with different data
	patterns, {\em iii)} test with timing-parameter combinations, {\em iv)} test
	with different temperatures, and {\em v)} DRAM read/write. Most of these
	scenarios show that a very high fraction (more than 95\%) of the erroneous
	cells consistently experience an error over multiple iterations of the same
	test (Section~7.6 of our HPCA 2015 paper~\cite{lee-hpca2015}).

	\item {\bf Performance Sensitivity Analyses.} We analyze the impact of
	increasing the number of ranks and channels, executing heterogeneous
	workloads, using different row buffer policies. We show that AL-DRAM
	effectively improves performance in all cases (Section~8.4
	of our HPCA 2015 paper~\cite{lee-hpca2015}).

	\item {\bf Power Consumption Analysis.} We show that AL-DRAM reduces DRAM
	power consumption by 5.8\%. This reduced power consumption is due to the reduced DRAM
	latencies (Section~8.4 of our HPCA 2015 paper~\cite{lee-hpca2015}).

\squishend

%% file: fig/worst.tex
\begin{figure}[h]
	\center
	\includegraphics[width=0.98\linewidth]{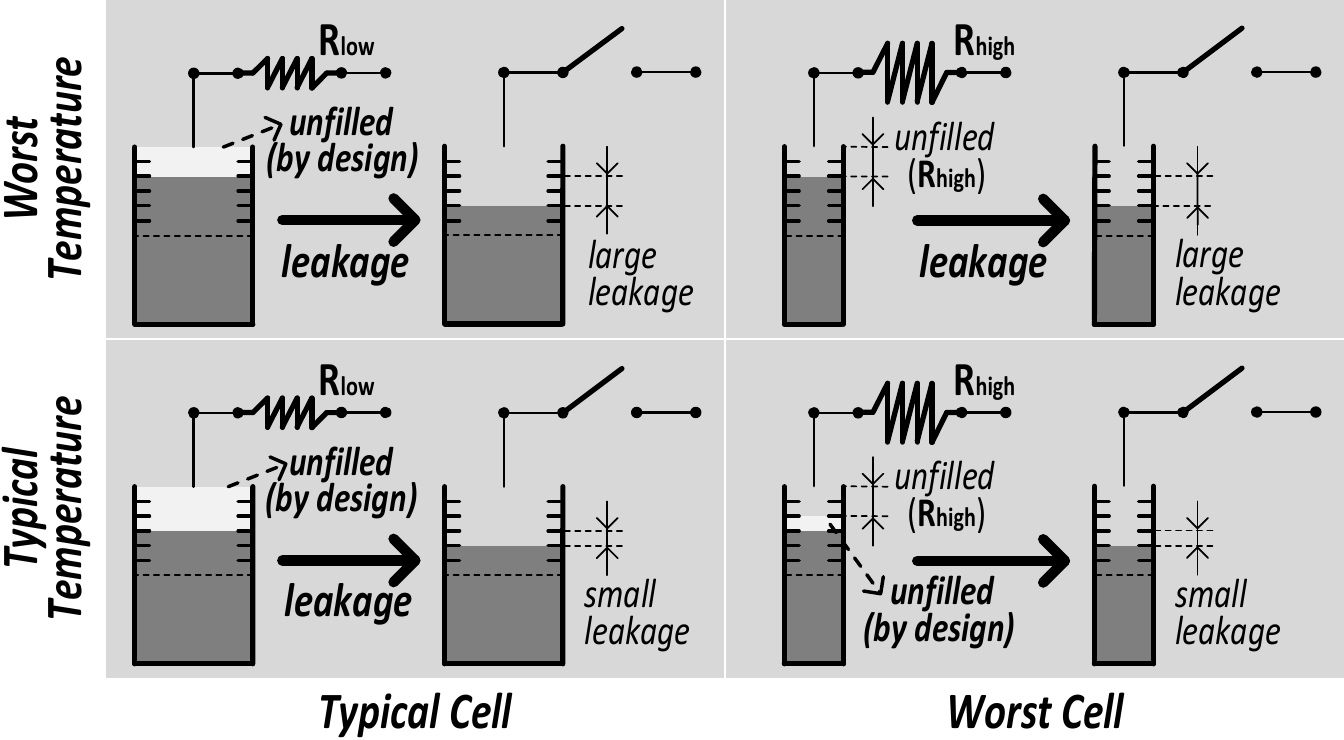}
	\caption{Effect of reduced latency: typical~vs.~worst-case. Reproduced from~\cite{lee-hpca2015}.}
	\label{fig:aldram}
\end{figure}

%% file: fig/profile_detail.tex
\begin{figure}[!t]
	\centering

	\subfloat [Maximum error-free refresh interval at 85\celsius\xspace (bank/chip/module)] {
		\includegraphics[width=0.98\linewidth]{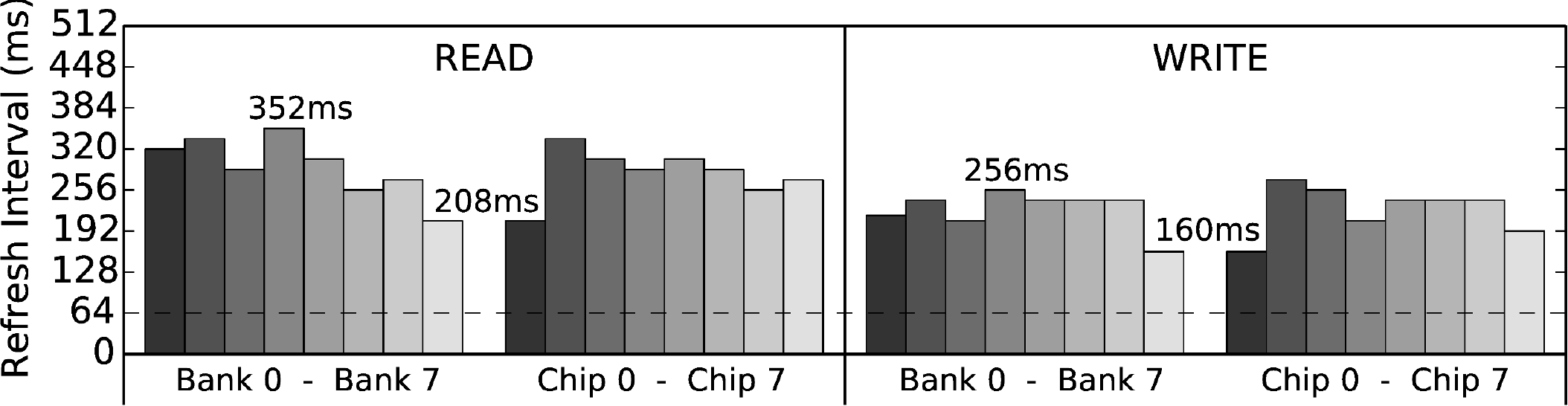}
		\label{fig:safety_detail}
	}

	\subfloat [Read latency (refresh Iinterval: 200 ms)] {
		\includegraphics[width=0.98\linewidth]{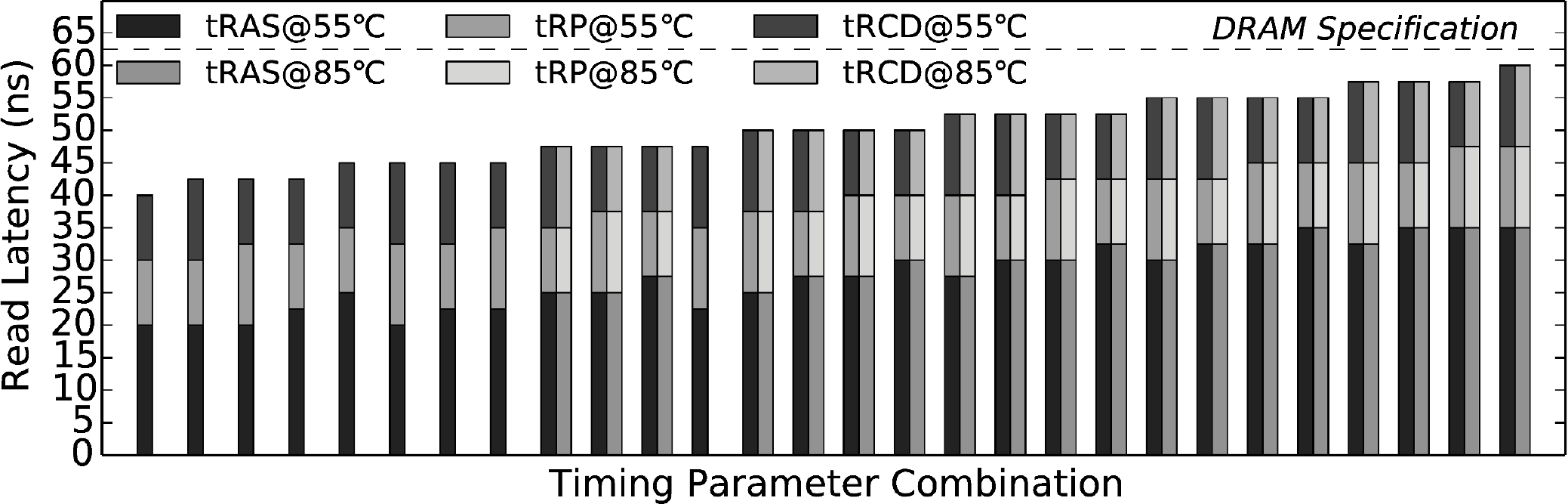}
		\label{fig:latency_detail_read}
	}

	\subfloat [Write latency (refresh interval: 152 ms)] {
		\includegraphics[width=0.98\linewidth]{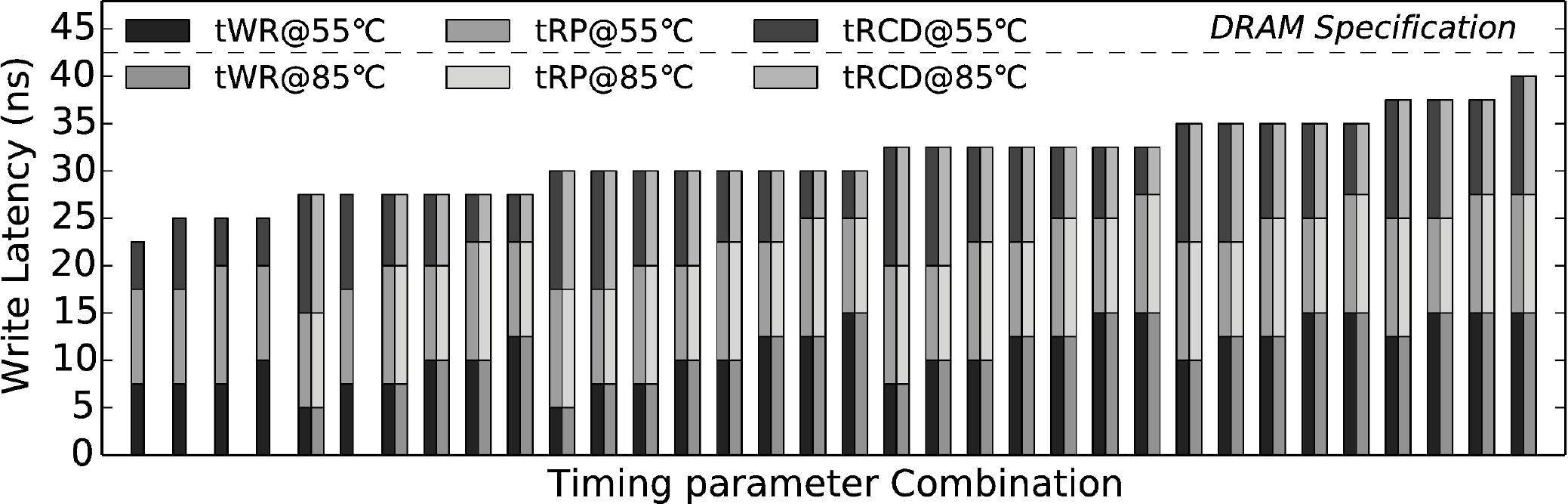}
		\label{fig:latency_detail_write}
	}
	\caption{Latency reductions while maintaining the safety margin of DRAM. Reproduced from~\cite{lee-hpca2015}.} \label{fig:latency_detail}
\end{figure}

%% file: fig/profile_multiple.tex
\begin{figure}[h]
	\centering
	\hspace{-0.05in}
	\subfloat [Read retention time] {
		\includegraphics[width=1.6in]{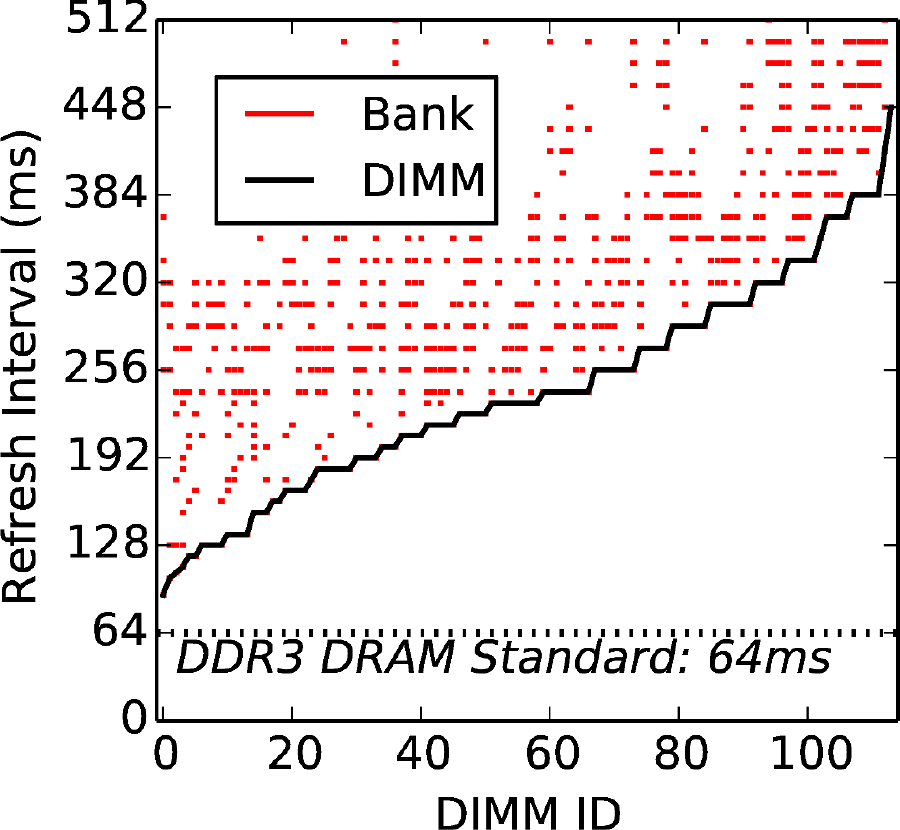}
		\label{fig:read_safety}
	}
	\subfloat [Write retention time] {
		\includegraphics[width=1.6in]{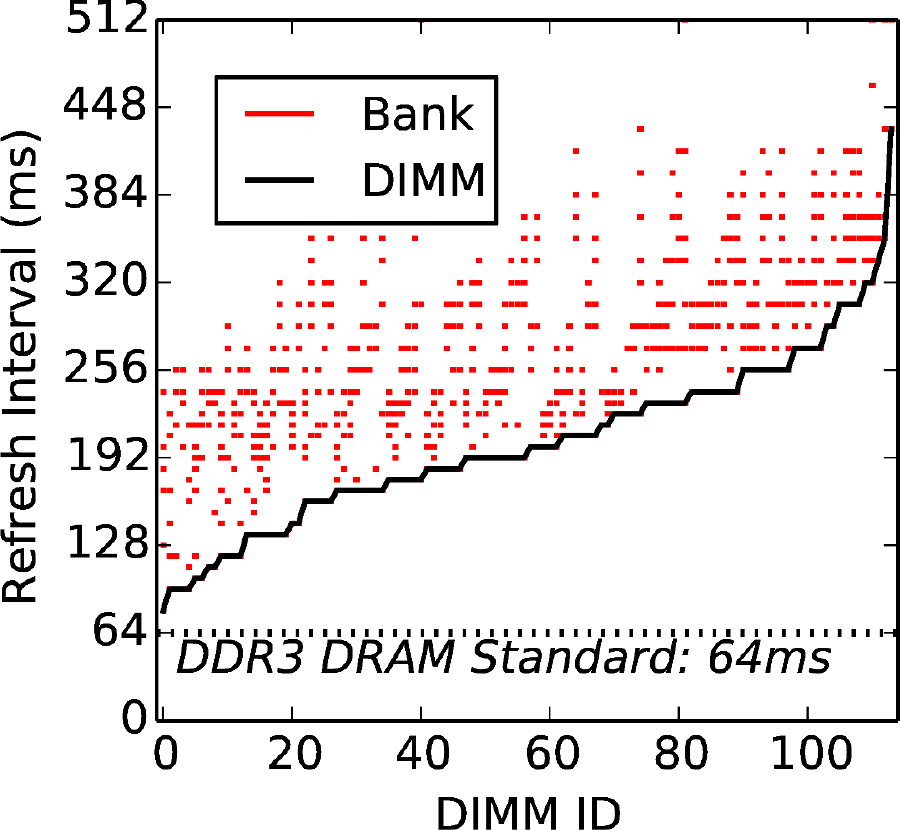}
		\label{fig:write_safety}
	}

	\hspace{-0.05in}
	\subfloat [Read latency] {
		\includegraphics[width=1.6in]{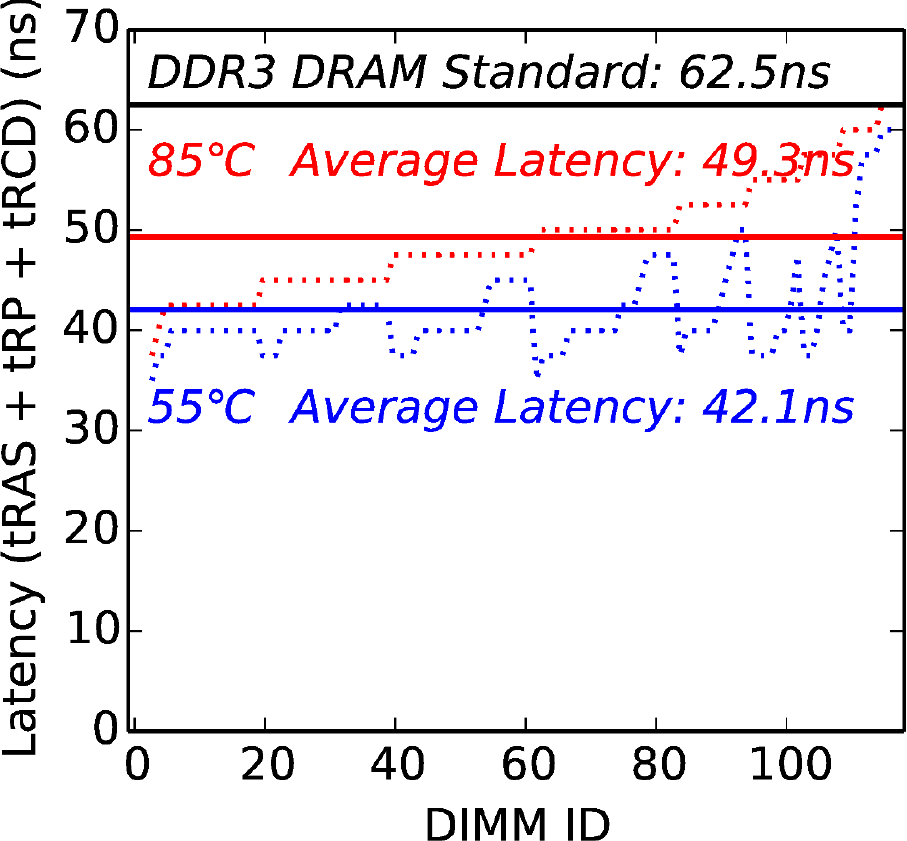}
		\label{fig:read_latency}
	}
	\subfloat [Write latency] {
		\includegraphics[width=1.6in]{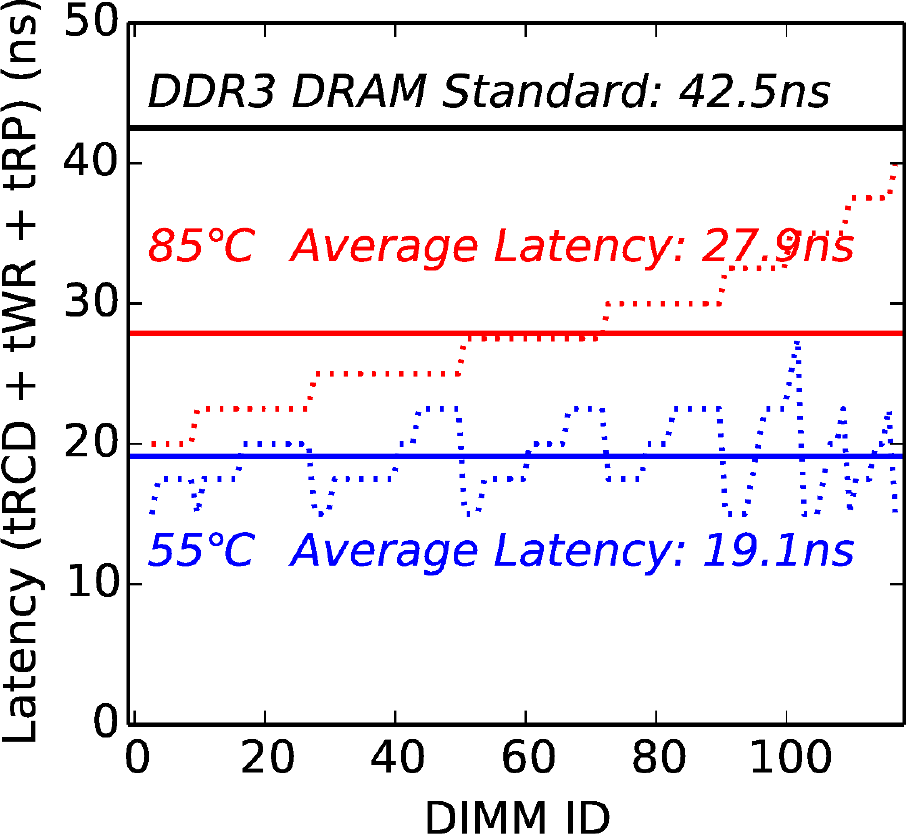}
		\label{fig:write_latency}
	}
	\caption{Analysis of 115 modules. Reproduced from~\cite{lee-hpca2015}.} \label{fig:multiple_dimms}
\end{figure}

%% file: fig/result.tex
\begin{figure}[h]
	\centering
	\includegraphics[width=0.99\linewidth]{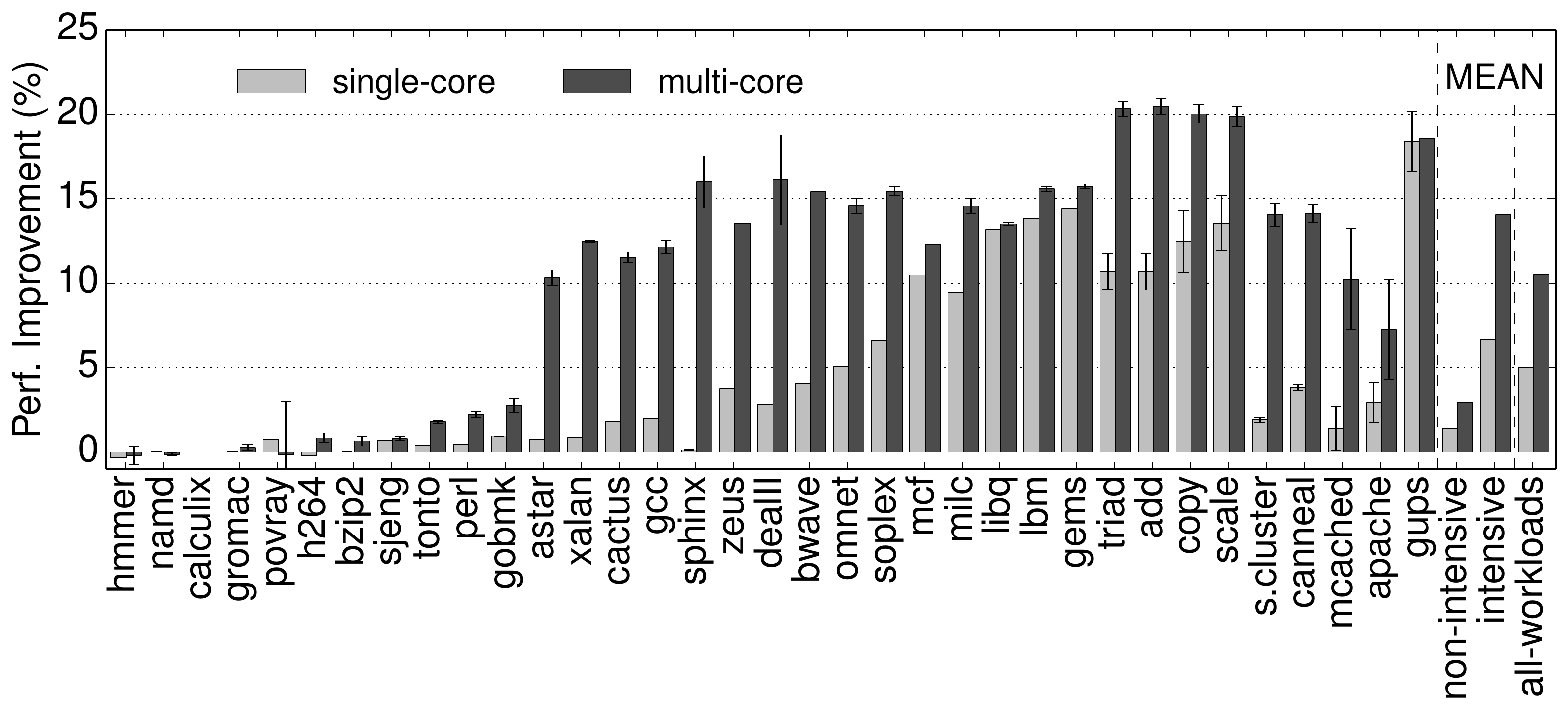}
	\captionof{figure}{Real system performance improvement with AL-DRAM. Reproduced from \cite{lee-hpca2015}.}
	\label{fig:result_1r1c}
\end{figure}

%% file: sec/1a_relatedwork.tex
\section{Related Work}
\label{sec:related}

To our knowledge, our HPCA 2015 paper is the first work to {\em i)} provide a
detailed qualitative and empirical analysis of the relationship between {\em
process variation} and {\em temperature dependence} of modern DRAM devices on
the one side, and DRAM access latency on the other side (we directly attribute
the relationship between the two to {\em the amount of charge} in cells), {\em
ii)} experimentally characterize a large number of existing DIMMs to understand
the potential of reducing DRAM timing constraints, {\em iii)} provide a
practical mechanism that can take advantage of this potential, and {\em iv)}
evaluate the performance benefits of this mechanism by {\em dynamically
optimizing} DRAM timing parameters on a real system using a variety of real
workloads.

Several works investigated the possibility of reducing DRAM latency by either
exploiting DRAM latency variation or changing the
DRAM architecture. We
discuss these below.

{\bf DRAM Latency Variation.} 
Chandrasekar et
al.~\cite{chandrasekar-date2014} evaluate the potential of relaxing some DRAM timing
parameters to reduce DRAM latency. This work observes latency variations across
DIMMs as well as for a DIMM at different operating temperatures. However, there
is no explanation as to why this phenomenon exists. In contrast, our HPCA 2015 paper~\cite{lee-hpca2015} {\em
(i)}~identifies and analyzes the root cause of latency variation in detail,
{\em (ii)}~provides a practical mechanism that can relax timing parameters, and
{\em (iii)}~provides a real system evaluation of this new mechanism, using real
workloads, showing improved performance and preserved reliability.

NUAT~\cite{shin-hpca2014} and ChargeCache~\cite{hassan-hpca2016} show that recently-refreshed
rows contain more charge, and propose mechanisms to access recently-refreshed
rows with reduced latency.
Even though some of the
observations in these works are similar to ours, the approaches to leverage
them are different. AL-DRAM exploits temperature dependence in a DIMM and
process variations across DIMMs, while NUAT and ChargeCache use the time difference between a row
refresh and an access to the row (hence its benefits are dependent on when the
row is accessed after it is refreshed). 
Therefore, NUAT and ChargeCache are complementary to AL-DRAM, and can
potentially be combined for better performance. 

Voltron~\cite{chang-sigmetrics2017} uses an experimental characterization
of real DRAM modules to identify the relationship between the DRAM supply voltage
and access latency variation.  Voltron uses this relationship to identify the
combination of voltage and access latency that minimizes system-level
energy consumption without exceeding a user-specified threshold for
the maximum acceptable performance loss.

Flexible-Latency DRAM (FLY-DRAM)~\cite{chang-sigmetrics2016} uses an experimental characterization
of real DRAM modules to capture access latency
variation across DRAM cells \emph{within} a single DRAM chip due to manufacturing
process variation.  FLY-DRAM identifies that there is spatial locality in the slower cells, resulting in \emph{fast regions} (i.e.,
regions where all DRAM cells can operate at significantly-reduced access 
latency without experiencing errors) and \emph{slow regions} (i.e., regions where \emph{some} of the DRAM
cells \emph{cannot} operate at significantly-reduced access latency without experiencing errors)
within each chip.
To take advantage of this heterogeneity in the reliable access latency of
DRAM cells within a chip,
FLY-DRAM (1)~categorizes the cells into fast and slow regions; and
(2)~lowers the overall DRAM latency by accessing fast regions with a lower 
latency.

Design-Induced Variation-Aware DRAM (DIVA-DRAM)~\cite{lee-sigmetrics2017} uses an experimental characterization
of real DRAM modules to identify the
latency variation within a single DRAM chip that occurs due to the architectural
design of the chip.  For example, a cell that is further away from the
row decoder requires a longer access time than a cell that is close to the row
decoder.  Similarly, a cell that is farther away from the wordline driver
requires a larger access time than a cell that is close to the wordline driver.
DIVA-DRAM uses design-induced variation to reduce the access latency
to different parts of the chip.
% Shin et al.~\cite{nuat} show that DRAM leakage affects two DRAM timing
% parameters ({\tt tRCD}/{\tt tRAS}), and propose a mechanism (NUAT) to access
% recently-refreshed rows with reduced latency, leveraging the fact that
% recently-refreshed rows have higher amount of charge. Even though some of the
% observations in~\cite{nuat} are similar to ours, the approaches to leverage
% them are different. AL-DRAM exploits temperature dependence in a DIMM and
% process variations across DIMMs. NUAT uses the time difference between a row
% refresh and an access to the row (hence its benefits are dependent on when the
% row is accessed after it is refreshed). AL-DRAM provides latency benefits
% regardless of access pattern and NUAT provides latency benefits regardless of
% DRAM temperature. Therefore, the two techniques are complementary and can be
% potentially combined for better performance. We note that {\em (i)} \ALD has
% very little overhead compared to NUAT, and {\em (ii)} our work experimentally
% analyzes the dependence of DRAM latency on temperature and process variation in
% {\em real} DRAM chips.

{\bf Low-Latency DRAM Architectures.} Various works~\cite{mutlu-memcon2013,
sato-vlsi1998, rldram, lee-hpca2013, son-isca2013, kim-isca2012,
seshadri-micro2013, hidaka-ieeemicro1990, zhang-ieeemicro2001, chang-hpca2014,
chang-hpca2016, chang-thesis2017} propose new DRAM architectures that provide
lower latency. Many of these works improve DRAM latency at the cost of either
significant additional DRAM chip area (i.e., extra sense
amplifiers~\cite{sato-vlsi1998, rldram, son-isca2013}, an additional SRAM
cache~\cite{hidaka-ieeemicro1990, zhang-ieeemicro2001}), specialized
protocols~\cite{seshadri-micro2013, kim-isca2012, lee-hpca2013, chang-hpca2014}
or a combination of these. Our proposed mechanism requires {\em no changes} to
the DRAM chip and the DRAM interface, and hence has almost negligible overhead.
Furthermore, \ALD is largely orthogonal to these proposed designs, and can be
applied in conjunction with them, providing greater cumulative reduction in
latency.

{\bf Binning or Overclocking DRAM.} \ALD has multiple sets of DRAM timing
parameters for different temperatures and dynamically optimizes the timing
parameters at runtime. Therefore, \ALD is different from simple binning
(performed by manufacturers) or over-clocking (performed by
end-users; e.g., \cite{intel-xmp, nvidia-sli}) that are used to figure out the highest
{\em static} frequency or lowest {\em static} timing parameters for DIMMs.

{\bf Other Methods for Lowering Memory Latency.} There are many works that
reduce {\em overall memory access latency} by modifying DRAM, the
DRAM-controller interface, and DRAM controllers. These works enable more
parallelism and bandwidth~\cite{kim-isca2012, chang-hpca2014,
seshadri-micro2013, lee-taco2016, chang-hpca2016, zhang-isca2014, ahn-taco2012,
ahn-cal2009, ware-iccd2006, zheng-micro2008, seshadri-micro2017,
seshadri-cal2015, lee-pact2015}, reduce refresh counts~\cite{liu-isca2012,
liu-isca2013, khan-sigmetrics2014, khan-cal2016, venkatesan-hpca2006,
qureshi-dsn2015, khan-micro2017}, accelerate bulk
operations~\cite{seshadri-micro2013, seshadri-cal2015, seshadri-micro2015,
chang-hpca2016, seshadri-micro2017}, accelerate computation in the logic layer
of 3D-stacked DRAM~\cite{ahn-isca2015a, ahn-isca2015b, zhang-hpca2014,
guo-wondp2014, boroumand-cal2016, hsieh-iccd2016, hsieh-isca2016,
pattnaik-pact2016, liu-spaa2017, kim-apbc2018, boroumand.asplos18}, enable better communication
between the CPU and other devices through DRAM~\cite{lee-pact2015}, leverage DRAM
access patterns~\cite{shin-hpca2014, hassan-hpca2016}, reduce write-related latencies by better
designing DRAM and DRAM control policies~\cite{chatterjee-hpca2012,
lee-techreport2010, seshadri-isca2014}, reduce overall queuing latencies in DRAM
by better scheduling memory requests~\cite{rixner-isca2000, moscibroda-podc2008,
lee-micro2009, nesbit-micro2006, moscibroda-usenix2007,
ausavarungnirun-isca2012, ausavarungnirun-pact2015, zhao-micro2014,
mutlu-micro2007, mutlu-isca2008, ebrahimi-micro2011, kim-hpca2010,
kim-micro2010, das-hpca2013, muralidhara-micro2011, jog-sigmetrics2016,
subramanian-tpds2016, subramanian-iccd2014, ipek-isca2008, usui-taco2016,
subramanian-micro2015, subramanian-hpca2013, lee-micro2008, lee-tc2011,
li-cluster2017, pattnaik-pact2016, ghose-isca2013, kaseridis-micro2011, hur-micro2004, shao-hpca2007, mukundan-hpca2012}, employ
prefetching~\cite{srinath-hpca2007, patterson-sosp1995, nesbit-pact2004,
ebrahimi-hpca2009, ebrahimi-micro2009, ebrahimi-isca2011, dahlgren-tpds1995,
alameldeen-hpca2007, cao-sigmetrics1995, lee-micro2008, mutlu-hpca2003,
mutlu-ieeemicro2003, mutlu-isca2005, mutlu-micro2005, dundas-ics1997,
cooksey-asplos2002, mutlu-ieeemicro2006}, perform memory/cache
compression~\cite{pekhimenko-micro2013, pekhimenko-pact2012, shafiee-hpca2014,
zhang-asplos2000, wilson-atec1999, dusser-ics2009, douglis-usenix1993,
decastro-sbacpad2003, alameldeen-tech2004, alameldeen-isca2004, abali-ibm2001,
pekhimenko-hpca2015, pekhimenko-hpca2016, vijaykumar-isca2015,
pekhimenko-ieeecal2015}, or perform better caching~\cite{seshadri-pact2012,
khan-hpca2014, qureshi-isca2007, qureshi-isca2006, seshadri-taco2015}.
Our proposal is orthogonal to all of these approaches and can be
applied in conjunction with them to achieve higher latency and energy benefits.

% {\bf DRAM Temperature-Based Memory Control.} Previous works~\cite{lin09, liu11}
% propose a temperature-based control mechanism to improve power efficiency and
% temperature balance across the memory hierarchy. Lin et al.~\cite{lin09} put
% overheated DRAMs to idle or power-downed states to distribute temperature
% across all ranks. Liu et al.~\cite{liu11} change memory access rates based on
% temperatures of individual DRAM chips by optimizing cache replacement and page
% allocation policies. While these works focus on the important problem of heat
% distribution in DRAMs, our work (with its focus on DRAM latency) is orthogonal
% to them.

\textbf{Experimental Studies of DRAM Chips.} 
There are several studies that characterize
various errors in DRAM. Many of these works observe how specific factors affect
DRAM errors, analyzing the impact of
temperature~\cite{elsayed-sigmetrics2012} and hard
errors~\cite{hwang-asplos2012}. Other works have conducted studies of DRAM error
rates in the field, studying failures across a large sample
size~\cite{schroeder-sigmetrics2009, meza-dsn2015, li-usenixatc2010,
  sridharan-sc2012, sridharan-sc2013, sridharan-asplos2015}. There are also works that have
studied errors through controlled experiments, 
usually using FPGA-based DRAM testing infrastructures like SoftMC~\cite{hassan-hpca2017},
to investigate errors due to
retention time~\cite{liu-isca2012, khan-cal2016,khan-micro2017,liu-isca2013,
    khan-sigmetrics2014, khan-dsn2016, qureshi-dsn2015, patel-isca2017,
    hassan-hpca2017}, disturbance from neighboring DRAM
cells~\cite{mutlu-date2017,kim-isca2014,kim-thesis,jung-memsys2016}, latency variation across/within DRAM
chips~\cite{lee-thesis2016,chandrasekar-date2014, chang-sigmetrics2016,
  lee-sigmetrics2017,chang-thesis2017}, and supply voltage~\cite{chang-sigmetrics2017,chang-thesis2017}. None of
these works extensively study latency variation across
DRAM modules, which we characterize in our work.

%% file: sec/2_significance.tex
\section{Significance} \label{sec:significance}

Our work on AL-DRAM is the first to extensively characterize and exploit
the large access latency variation that exists in modern DRAM devices.
In this section, we discuss the novelty of AL-DRAM and its expected
future impact on the community.

\subsection{Novelty} \label{sec:novelty}

We make the following major contributions in our HPCA 2015 paper~\cite{lee-hpca2015}:

{\bf Addressing a Critical Real Problem, High DRAM Latency, with Low Cost.} High
DRAM latency is a critical bottleneck for overall system performance in a
variety of modern computing systems~\cite{mutlu-superfri2015, mutlu-imw2013},
especially in real large-scale server systems~\cite{lo-isca2015,
kanev-isca2015}. Considering the significant difficulties in DRAM
scaling~\cite{mutlu-superfri2015, mutlu-imw2013, kang-memforum2014,
mutlu-date2017}, the DRAM latency problem is getting worse in future systems due to process
variation. Our HPCA 2015 work~\cite{lee-hpca2015} leverages the heterogeneity created by DRAM
process variation across DRAM chips and system operating conditions to mitigate
the DRAM latency problem. We propose a practical mechanism, {\em
Adaptive-Latency DRAM}, which mitigates DRAM latency with very modest hardware
cost, and with {\em no changes} to the DRAM chip itself.

% {\bf Low-Latency DRAM Architectures.} Various works~\cite{mutlu-memcon2013,
% sato-vlsi1998, rldram, lee-hpca2013, son-isca2013, kim-isca2012,
% seshadri-micro2013, hidaka-ieeemicro1990, zhang-ieeemicro2001, chang-hpca2014,
% chang-hpca2016, chang-thesis2017} propose new DRAM architectures that provide
% lower latency. Many of these works improve DRAM latency at the cost of either
% significant additional DRAM chip area (i.e., extra sense
% amplifiers~\cite{sato-vlsi1998, rldram, son-isca2013}, an additional SRAM
% cache~\cite{hidaka-ieeemicro1990, zhang-ieeemicro2001}), specialized
% protocols~\cite{seshadri-micro2013, kim-isca2012, lee-hpca2013, chang-hpca2014}
% or a combination of these. Our proposed mechanism requires {\em no changes} to
% the DRAM chip and the DRAM interface, and hence has almost negligible overhead.
% Furthermore, \ALD is largely orthogonal to these proposed designs, and can be
% applied in conjunction with them, providing greater cumulative reduction in
% latency.

{\bf Large-Scale Latency Profiling of Modern DRAM Chips.} Using our FPGA-based
DRAM testing infrastructure~\cite{liu-isca2013, khan-sigmetrics2014,
kim-isca2014, qureshi-dsn2015, lee-hpca2015, khan-dsn2016, chang-sigmetrics2016,
chang-sigmetrics2017, lee-thesis2016, chang-thesis2017, hassan-hpca2017,
lee-sigmetrics2017, patel-isca2017}, we profile 115 DRAM modules
(920 DRAM chips in total) and show that there is significant timing variation
between different DIMMs at different temperatures. We believe that our results
are statistically significant to validate our hypothesis that the DRAM timing
parameters strongly depend on the amount of cell charge. We provide a detailed
characterization of each DIMM online at the SAFARI Research Group
website~\cite{safari}. Furthermore, we introduce our FPGA-based DRAM
infrastructure and experimental methodology for DRAM profiling, which are
carefully constructed to represent the worst-case conditions in power noise,
bitline/wordline coupling, data patterns, and access patterns. Such information
will hopefully be useful for future DRAM research.

{\bf Extensive {\em Real} System Evaluation of DRAM Latency.} We evaluate our
mechanism on a real system~\cite{amd-4386, amd-bkdg} and show that our mechanism
provides significant performance improvements. Reducing the timing parameters
strips the excessive margin in the electrical charge stored within a DRAM cell. We show that the
remaining margin is {\em enough} for DRAM to operate reliably. To verify the
correctness of our experiments, we ran our workloads for 33~days nonstop, and
examined their and the system's correctness with reduced timing parameters.
Using the reduced timing parameters over the course of 33~days, our real system
was able to execute 35 different workloads in both single-core and multi-core
configurations while preserving correctness and being {\em error-free}. Note
that these results do {\em not} absolutely guarantee that no errors can be
introduced by reducing the timing parameters. However, we believe that we have
demonstrated a proof-of-concept which shows that DRAM latency can be reduced at
no impact on DRAM reliability. Ultimately, DRAM manufacturers can provide
the reliable timing parameters for different operating conditions and modules.

\subsection{Potential Long-Term Impact} \label{sec:longterm}

{\bf Tolerating High DRAM Latency by Exploiting DRAM Intrinsic Characteristics.}
Today, there is a large latency cliff between the on-chip last level cache and
off-chip DRAM, leading to a large performance fall-off when applications start
missing in the last level cache. By enabling lower DRAM latency, our mechanism,
Adaptive-Latency DRAM, smoothens this latency cliff without adding another layer
into the memory hierarchy.

{\bf Applicability to Future Memory Devices.} We show the benefits of the
common-case timing optimization in modern DRAM devices by taking advantage of
intrinsic characteristics of DRAM. Considering that most memory devices adopt a
unified specification that is dictated by the worst-case operating condition,
our approach that optimizes device latency for the common case can be applicable
to other memory devices by leveraging the intrinsic characteristics of the
technology they are built with. We believe there is significant potential for
approaches that could reduce the latency of Phase Change Memory
(PCM)~\cite{raoux-ibm2008, lee-isca2009, qureshi-isca2009, qureshi-micro2009,
dhiman-dac2009, lee-ieeemicro2010, lee-cacm2010, meza-iccd2012, yoon-taco2014,
wong-ieee2010}, STT-MRAM~\cite{kultursay-ispass2013, meza-iccd2012}, 
RRAM~\cite{wong-ieee2012}, and NAND flash memory~\cite{luo-msst2015,
cai-date2013, cai-itj2013, cai-iccd2012, cai-sigmetrics2014, cai-iccd2013,
cai-dsn2015, cai-hpca2015, meza-sigmetrics2015, lu-tc2015, cai-date2012,
cai-dsn2015, cai-hpca2017, luo-jsac2016, cai-ieee2017, cai.procieee.arxiv17,
cai.bookchapter.arxiv17}.

{\bf New Research Opportunities.} Adaptive-Latency DRAM creates new
opportunities by enabling mechanisms that can leverage the heterogeneous latency
offered by our mechanism. We describe a few of these briefly.

{\em Optimizing the operating conditions for faster DRAM access:}
Adaptive-Latency DRAM provides different access latencies for different operating
conditions. Future works can explore how the operating conditions themselves
can be optimized, which can be used in conjunction with AL-DRAM to 
further improve the DRAM access latency.
For instance, balancing DRAM accesses over
multiple DRAM channels and ranks can potentially reduce the DRAM operating
temperature, maximizing the benefits provided by AL-DRAM. At the system
level, operating the system at a constant low temperature can enable the use of
lower DRAM latencies more frequently.

{\em Optimizing data placement to reduce overall DRAM access latency:} We
characterize the latency variation in different DIMMs due to process variation.
Placing data based on this information and the latency criticality of data
maximizes the benefits of lowering DRAM latency, by placing the data
that is most sensitive to latency in the fastest DRAM chips (and, thus,
providing lookups to the data with the fastest access latency).

{\em Error correction mechanisms to further reduce DRAM latency.}
Error correction mechanisms allow us to lower DRAM latency even
further, by correcting bit errors that occur when a small number of the
DRAM operations end before the minimum charge is stored in the DRAM cell.
Such mechanisms can rely on error correction to compensate for the
reduced reliability of read and write operations at even lower latencies, 
leading to a further reduction in DRAM latency without errors. Future
research that uses error correction to enable even lower latency DRAM is
therefore promising as it opens a new set of trade-offs.
Note that our recent work, DIVA-DRAM~\cite{lee-sigmetrics2017}, explores this
direction and finds very promising benefits.

\ch{Inspired by our characterization and proposed techniques,
several recent works~\cite{chang-sigmetrics2016, chang-sigmetrics2017, 
lee-sigmetrics2017, kim-hpca2018, hassan-hpca2016, patel-isca2017}
have explored many of these new research
opportunities, by (1)~analyzing
different sources of latency and performance variation within DRAM
chips, and 
(2)~exploiting these sources of latency and performance variation to
reduce access latency and/or energy consumption.}

%% file: sec/2a_conclusion.tex
\section{Conclusion}

This paper summarizes our HPCA 2015 work on Adaptive-Latency DRAM (AL-DRAM),
a simple and effective mechanism for dynamically tailoring the
DRAM timing parameters for the current operating condition
without introducing any errors.  
AL-DRAM takes advantage of the large latency margin available
in the DRAM timing parameters for common-case operation,
by dynamically the operating temperature
of each DRAM module and employing timing constraints optimized for
a particular module at the current temperature.
AL-DRAM provides 
an average 14\% improvement in overall system performance
across a wide variety of memory-intensive applications run on
a real multi-core system.
We  conclude  that  AL-DRAM  is  a  simple  and  effective
mechanism to reduce DRAM latency.   We hope that our experimental 
exposure of the large margin present in the standard
DRAM timing constraints will inspire other approaches to optimize 
DRAM chips, latencies, and parameters at low cost.

%% file: sec/3_acknowledge.tex
\section*{Acknowledgments}

We thank Saugata Ghose for his dedicated effort in the preparation
of this article.
We thank the anonymous reviewers for their valuable feedback. We thank Uksong
Kang, Jung-Bae Lee, and Joo Sun Choi from Samsung, and Michael Kozuch from Intel
for their helpful comments. We acknowledge the support of our industrial
partners: Facebook, IBM, Intel, Microsoft, Qualcomm, VMware, and Samsung. This
research was partially supported by NSF (grants 0953246, 1212962, 1065112),
the Semiconductor Research Corporation, and the Intel Science and Technology Center
for Cloud Computing. Donghyuk Lee was supported in part by the John and Claire
Bertucci Graduate Fellowship.